\documentclass{article}

\PassOptionsToPackage{numbers, compress}{natbib}

\usepackage[final]{neurips_2023}
\usepackage{graphicx}
\usepackage{tabularx}
\usepackage[utf8]{inputenc} 
\usepackage[T1]{fontenc}    
\usepackage{hyperref}       
\usepackage{url}            
\usepackage{booktabs}       
\usepackage{amsfonts}       
\usepackage{nicefrac}       
\usepackage{microtype}      
\usepackage{xcolor}         
\usepackage{multirow}       
\usepackage{array}          
\usepackage{enumitem}       

\title{How to design an AI ethics board}

\author{
  Jonas Schuett\thanks{Equal contribution} \\
  Centre for the Governance of AI \\
  \texttt{jonas.schuett@governance.ai} \\
  \And
  Anka Reuel$^*$ \\
  Stanford University \\
  \texttt{anka@cs.stanford.edu} \\
  \AND
  Alexis Carlier \\
  Centre for the Governance of AI \\
  \texttt{alexispcarlier@gmail.com} \\
}

\begin{document}

\maketitle

\begin{abstract}
  Organizations that develop and deploy artificial intelligence (AI) systems need to take measures to reduce the associated risks. In this paper, we examine how AI companies could design an AI ethics board in a way that reduces risks from AI. We identify five high-level design choices: (1) What responsibilities should the board have? (2) What should its legal structure be? (3) Who should sit on the board? (4) How should it make decisions and should its decisions be binding? (5) What resources does it need? We break down each of these questions into more specific sub-questions, list options, and discuss how different design choices affect the board’s ability to reduce risks from AI. Several failures have shown that designing an AI ethics board can be challenging. This paper provides a toolbox that can help AI companies to overcome these challenges.
\end{abstract}

\section{Introduction} \label{Ch:1:introduction}

It becomes increasingly clear that state-of-the-art artificial intelligence (AI) systems pose significant societal risks. AI systems used for drug discovery could be misused for the design of biochemical weapons \cite{urbina2022dual}. A failure of AI systems used to control nuclear power plants or other critical infrastructure could also have devastating consequences \cite{degrave2022magnetic}. Another concern is that, as models become larger and larger, certain dangerous capabilities might emerge at some point. Scholars and practitioners are increasingly worried about power-seeking behavior, situational awareness, and the ability to persuade people \cite{carlsmith2022powerseeking, ngo2023alignment, openai2023gpt4technical}. Organizations that develop and deploy AI systems need to take measures to reduce these risks to an acceptable level. In this paper, we examine how AI companies could design an AI ethics board in a way that reduces risks from AI. By “ethics board”, we mean a collective body intended to promote an organization’s ethical behavior.

Some AI companies already have an AI ethics board. For example, Meta’s Oversight Board makes binding decisions about the content on Facebook and Instagram \cite{oversightboard2023website, klonick2020facebook, wong2022meta}. Microsoft’s AI, Ethics and Effects in Engineering and Research (AETHER) Committee advises their leadership “on the challenges and opportunities presented by AI innovations” \cite{microsoft2023ourapproach}. DeepMind’s Institutional Review Committee (IRC) oversees their human rights policy \cite{deepmind2023humanrightspolicy} and has already played a key role in the AlphaFold release \cite{kavukcuoglu2022alphafold}. These examples show that AI ethics boards are of practical relevance.

But there have also been a number of failures. Google’s Advanced Technology External Advisory Council (ATEAC) faced significant resistance over the inclusion of disputable members. It was shut down only one week after its announcement \cite{piper2019googlefail, acquisti2019tweet, transphobia2019googlers, walker2019external}. Axon’s AI and Policing Technologies Ethics Board was effectively discontinued in June 2022 after three years of operations \cite{smith2022axon}. Nine out of eleven members resigned after Axon announced plans to develop taser-equipped drones to be used in schools without consulting the board first \cite{policing2022resigning}. (In late 2022, Axon announced their new ethics board: the Ethics \& Equity Advisory Council [EEAC], which gives feedback on a limited number of products “through a racial equity and ethics lens” \cite{axon2022ethicsequity}.) These cases show that designing an AI ethics board can be challenging. It also highlights the need for more research.

Although there has been some research on AI ethics boards, the topic remains understudied.
The most important work for our purposes is a whitepaper by Accenture \cite{sandler2019building}. They discuss key benefits of AI ethics boards and identify key design questions. However, their discussion lacks both breadth and depth. They discuss only a handful of design considerations and do not go into detail. They also do not focus on leading AI companies and risk reduction. Besides that, there is some literature on the purpose \cite{jordan2019designing, tiell2019create, morley2021ethics} and practical challenges of AI ethics boards \cite{adalovelace2022looking, gupta2020ai}. There are also several case studies of existing boards, including Meta’s Oversight Board \cite{wong2022meta} and Microsoft’s AETHER Committee \cite{newman2020decisionpoints}. And finally, there is some discussion of the role of AI ethics boards in academic research \cite{bernstein2021ethics, srikumar2022advancing}. Taken together, there seem to be at least two gaps in the literature. First, there is only limited work on the practical question of how to design an AI ethics board. Second, there is no discussion of how specific design considerations can help to reduce risks from AI. In light of these gaps, the paper seeks to answer two research questions (RQs):

\begin{itemize}[leftmargin=2em]
    \item \textbf{RQ1:} What are the key design choices that AI companies have to make when setting up an AI ethics board?
    \item \textbf{RQ2:} How could different design choices affect the board’s ability to reduce risks from AI?
\end{itemize}

The paper has two areas of focus. First, it focuses on organizations that develop state-of-the-art AI systems. This includes medium-sized research labs (e.g. OpenAI, DeepMind, and Anthropic) and big tech companies (e.g. Microsoft and Google). We use the term “AI company” or “company” to refer to them. 
Although we do not mention other types of companies (e.g. hardware companies), we expect that they might also benefit from our analysis. Second, the paper focuses on the board’s ability to reduce risks (see RQ2). By “risk”, we mean the “combination of the probability of occurrence of harm and the severity of that harm” \cite{iso51_2014}. (But note that there are other risk definitions \cite{iso31000_2018}). In terms of severity, we focus on adverse effects on large groups of people and society as a whole, especially threats to their lives and physical integrity. We are less interested in financial losses and risks to organizations themselves (e.g. litigation or reputation risks). In terms of likelihood, we also consider low-probability, high-impact risks, sometimes referred to as “black swans” \cite{taleb2007black, aven2013meaning, kolt2023algorithmic}. The two main sources of harm (“hazards”) we consider are accidents \cite{amodei2016concrete, arnold2021aiaccidents} and cases of misuse \cite{brundage2018malicious, goldstein2023generative, anderljung2023protecting}.

In the following, we consider five high-level design choices: What responsibilities should the board have (Section \ref{Ch:2:responsibilities})? What should its legal structure be (Section \ref{Ch:3:structure})? Who should sit on the board (Section \ref{Ch:4:membership})? How should it make decisions and should its decisions be binding (Section \ref{Ch:5:decision-making})? What resources does it need (Section \ref{Ch:6:resources})? We break down each of these questions into more specific sub-questions, list options, and discuss how they could affect the board’s ability to reduce risks from AI. The paper concludes with a summary of the most important design considerations and suggestions for further research (Section \ref{Ch:7:conclusion}).

\section{Responsibilities}\label{Ch:2:responsibilities}

What responsibilities should the board have? We use the term “responsibility” to refer to the board’s purpose (what it aims to achieve), its rights (what it can do), and duties (what it must do). The board’s responsibilities are typically specified in its charter or bylaws. In the following, we focus on responsibilities that could help to reduce risks from AI (see RQ2). The ethics board could advise the board of directors (Section \ref{Ch:2:1:advising}), oversee model releases and publications (Section \ref{Ch:2:2:overseeing}), support risk assessments (Section \ref{Ch:2:3:supporting}), review the company’s risk management practices (Section \ref{Ch:2:4:reviewving}), interpret AI ethics principles (Section \ref{Ch:2:5:interpreting}), or serve as a contact point for whistleblowers (Section \ref{Ch:2:6:contact}). Note that these responsibilities are neither mutually exclusive nor collectively exhaustive. The board could also have more than one responsibility.

\subsection{Advising the board of directors}\label{Ch:2:1:advising}

The board of directors plays a key role in the corporate governance of AI companies \cite{cihon2021corporate}. It sets the company’s strategic priorities, is responsible for risk oversight, and has significant influence over management (e.g. it can replace senior executives). But since many board members only work part-time and rely on information provided to them by management, they need support from an independent ally in the company \cite{davies2018three}. Internal audit can be this ally, but the ethics board could serve as an additional layer of assurance \cite{schuett2022lines}.

\paragraph{Options.} The ethics board could provide strategic advice on various topics. It could advocate against high-risk decisions and call for a more prudent and wiser course.

\begin{itemize}[leftmargin=2em]
    \item \textbf{Research priorities.} Most AI companies have an overarching research agenda (e.g. DeepMind’s focus on reinforcement learning \cite{silver2021reward} or Anthropic’s focus on empirical safety research \cite{anthropic2023coreviews}). This agenda influences what projects the company works on. The ethics board could try to influence that agenda. It could advocate for increasing focus on safety and alignment research \cite{amodei2016concrete, hendrycks2022unsolved, ngo2023alignment}. More generally, it could caution against advancing capabilities faster than safety measures. The underlying principle is called “differential technological development” \cite{bostrom2001analyzing, ord2020precipice, sandbrink2022differential}.
    \item \textbf{Commercialization strategy.} The ethics board could also advise on the company’s commercialization strategy. On the one hand, it is understandable that AI companies want to monetize their systems (e.g. to pay increasing costs for compute \cite{sevilla2022compute}). On the other hand, commercial pressure might incentivize companies to cut corners on safety \cite{armstrong2016racing, naude2020race}. For example, Google famously announced to “recalibrate” the level of risk it is willing to take in response to OpenAI’s release of ChatGPT \cite{grant2023googlehelp}. It has also been reported that disagreements over OpenAI’s commercialization strategy were the reason why key employees left the company to start Anthropic \cite{waters2021rebelai}.
    \item \textbf{Strategic partnerships.} AI labs might enter into strategic partnerships with profit-oriented companies (see e.g. the extended partnership between Microsoft and OpenAI \cite{microsoft2023openai}) or with the military (see e.g. “Project Maven”, Google’s collaboration with the U.S. Department of Defense \cite{conger2018pentagonai}). Although such partnerships are not inherently bad, they could contribute to an increase of risk (e.g. if they lead to an equipment of nuclear weapons with AI technology \cite{maas2019viable}).
    \item \textbf{Fundraising and M\&A transactions.} AI companies frequently need to bring in new investors. For example, in January 2023, it has been reported that OpenAI raised \$10B from Microsoft \cite{hoffman2023microsoftchatgpt, openai2023microsoft}. But if new investors care more about profits, this could gradually shift the company’s focus away from safety and ethics towards profit maximization. The same might happen if AI companies merge or get acquired. The underlying phenomena is called “mission drift” \cite{grimes2019anchors}.
\end{itemize}

\paragraph{Discussion.} How much would advising the board of directors reduce risk? This depends on many different factors. It would be easier if the ethics board has a direct communication channel to the board of directors, ideally to a dedicated risk committee. It would also be easier if the board of directors is able to do something about risks. They need risk-related expertise and governance structures to exercise their power (e.g. a chief risk officer [CRO] as a single point of accountability). But the board of directors also needs to take risks seriously and be willing to do something about them. This will often require a good relationship between the ethics board and the board of directors. Inversely, it would be harder for the ethics board to reduce risk if the board of directors mainly cares about other things (e.g. profits or prestige), especially since the ethics board is usually not able to force the board of directors to do something.

\subsection{Overseeing model releases and publications}\label{Ch:2:2:overseeing}

Many risks are caused by accidents \cite{amodei2016concrete, arnold2021aiaccidents} or the misuse of specific AI systems \cite{brundage2018malicious, goldstein2023generative, anderljung2023protecting}. In both cases, the deployment decision is a decisive moment. Ideally, companies should discover potential failure modes and vulnerabilities before they deploy a system, and stop the deployment process if they cannot reduce risks to an acceptable level. But not all risks are caused by the deployment of individual models. Some risks also stem from the publication of research, as research findings can be misused \cite{urbina2022dual, brundage2018malicious, goldstein2023generative, anderljung2023protecting, ashurst2022ai, shevlane2020offense, bostrom2019vulnerable}. The dissemination of potentially harmful information, including research findings, is called “infohazards” \cite{bostrom2011information, leahy2022conjecture}. Publications can also fuel harmful narratives. For example, it has been argued that the “arms race” rhetoric is highly problematic \cite{cave2018ai}.

\paragraph{Options.} An ethics board could try to reduce these risks by creating a release strategy \cite{solaiman2019release, solaiman2023gradient, openai2022bestpractices} and norms for the responsible publication of research \cite{crootof2019artificial, ashurst2022ai, shevlane2022structured, partnershiponai2021managingrisks}. For example, the release strategy could establish “structured access” as the norm for deploying powerful AI systems \cite{shevlane2022structured}. Instead of open-sourcing new models, companies might want to deploy them via an application programming interface (API), which would allow them to conduct know-your-customer (KYC) screenings and restrict access if necessary, while allowing the world to use and study the model. The release strategy could also specify instances where a “staged release” seems adequate. Stage release refers to the strategy of releasing a smaller model first, and only releasing larger models if no meaningful cases of misuse are observed. OpenAI has coined the term and championed the approach when releasing GPT-2 \cite{solaiman2019release}. But note that the approach has also been criticized \cite{crootof2019artificial}. The ethics board could also create an infohazard policy. The AI research organization Conjecture has published its policy \cite{leahy2022conjecture}. We expect most AI companies to have similar policies, but do not make them public. In addition to that, the board could oversee specific model releases and publications (not just the abstract strategies and policies). It could serve as an institutional review board (IRB) that cares about safety and ethics more generally, not just the protection of human subjects \cite{bernstein2021ethics, srikumar2022advancing}. In particular, it could review the risks of a model or publication itself, do a sanity check of existing reviews, or commission an external review (Section \ref{Ch:2:3:supporting}).

\paragraph{Discussion.} How much would this reduce risk? Among other things, this depends on whether board members have the necessary expertise (Section \ref{Ch:4:4characteristics:}), whether the board’s decisions are binding (Section \ref{Ch:5:2:binding}), and whether they have the necessary resources (Section \ref{Ch:6:resources}). The decision to release a model or publish research is one of the most important points of intervention for governance mechanisms that are intended to reduce risks. An additional attempt to steer such decisions in a good direction therefore seems desirable.

\subsection{Supporting risk assessments}\label{Ch:2:3:supporting}

By “risk assessment”, we mean the identification, analysis, and evaluation of risks \cite{iso51_2014, iso31000_2018}. Assessing the risks of state-of-the-art AI systems is extremely difficult: (1) The risk landscape is highly complex and evolves rapidly. For example, the increasing use of so-called “foundation models” \cite{bommasani2022opportunities} might lead to new diffuse and systemic risks (e.g. threats to epistemic security \cite{seger2022defence}). (2) Defining normative thresholds is extremely difficult: What level of risk is acceptable? How fair is fair enough? (3) In many cases, AI companies are also detached from the people who are most affected by their systems, often historically marginalized communities \cite{mohamed2020decolonial, birhane2022power}. (4) Risk assessments might become even more difficult in the future. For example, systems might become capable of deceiving their operators and only “pretending” to be safe in a testing environment \cite{ngo2023alignment}.

\paragraph{Options.} The ethics board could actively contribute to the different steps of a risk assessment. It could use a risk taxonomy to flag missing hazards \cite{weidinger2021ethical}, comment on a heatmap that illustrates the likelihood and severity of a risk \cite{iec31010_2019}, or try to circumvent a safety filter \cite{rando2022redteaming}. It could also commission a third-party audit \cite{raji2019actionable, brundage2020trustworthy, falco2021governing, raji2022outsider, mokander2021ethics, mökander2023auditing} or red team \cite{ganguli2022red, perez2022red, rando2022redteaming}. It could report its findings to the board of directors which would have the necessary power to intervene (Section \ref{Ch:2:1:advising}). Depending on its power, it might even be able to veto or at least delay deployment decisions (Section \ref{Ch:5:2:binding}).

\paragraph{Discussion.} Some companies already take extensive measures to assess risks before deploying state-of-the-art AI systems \cite{kavukcuoglu2022alphafold, openai2022lessonslearned, anthropic2023coreviews}. It is unclear how much value the support of an ethics board would add to such efforts. But especially when dealing with catastrophic risks, having an additional “layer of defense” seems generally desirable. The underlying concept is called “defense in depth” \cite{cotton2020defence}. This approach could be seen as a solution to the problem that “there is no silver bullet” 
\cite{openai2022lessonslearned}. But supporting risk assessments could also have negative effects. If other teams rely on the board’s work, they might assess risks less thoroughly. This would be particularly problematic if the board is not able to do it properly (e.g. it can only perform sanity checks). But this effect could be mitigated by clearly communicating expectations and creating appropriate incentives.

\subsection{Reviewing risk management practices}\label{Ch:2:4:reviewving}

Instead of or in addition to supporting specific risk assessments (Section \ref{Ch:2:3:supporting}), the ethics board could review the company’s risk management practices more generally. In other words, it could try to improve the company’s “risk governance” \cite{van2011risk, lundqvist2015firms}. Risk management practices at AI companies seem to be less advanced compared to other industries like aviation \cite{hunt2020flight}. “They might look good on paper, but do not work in practice” \cite{schuett2022lines}. There are not yet any established best practices and companies rarely adhere to best practices from other industries (though there are promising developments around risk management standards). And practices that companies develop themselves might not be as effective. For example, there might be blind spots for certain types of risks (e.g. diffuse or systemic risks) or they might not account for cognitive biases (e.g. availability bias or scope neglect \cite{yudkowsky2008cognitive}).

\paragraph{Options.} The ethics board could assess the adequacy and effectiveness of the company’s risk management practices. It could assess whether the company complies with relevant regulations \cite{schuett2022risk}, standards \cite{nist2023artificial, iso23894_2023}, or its own policies and processes. It could also try to find flaws in a more open-ended fashion. Depending on its expertise and capacity, it could do this on its own (e.g. by reviewing risk-related policies and interviewing people in risk-related positions) or commission an external review of risk management practices (e.g. by an audit firm \cite{mokander2022operationalising}). Note that this role is usually performed by the company’s internal audit function, but the ethics board could provide an additional layer of assurance \cite{schuett2022lines}. They could report their findings directly to the risk committee of the board of directors and the chief risk officer (CRO) who could make risk management practices more effective.

\paragraph{Discussion.} If companies already have an internal audit function, the additional value would be limited; the ethics board would merely be an additional defense layer \cite{schuett2022lines}. However, if companies do not already have an internal audit function, the added value could be significant. Without a deliberate attempt to identify ineffective risk management practices, some limitations will likely remain unnoticed \cite{schuett2022lines}. But the value ultimately depends on the individuals who conduct the review. This might be problematic because it will require a very specific type of expertise that most members of an ethics board do not have (Section \ref{Ch:4:4characteristics:}). It is also very time-consuming, so a part-time board might not be able to do it properly (Section \ref{Ch:4:5:time}). Both issues should be taken into account when appointing members.

\subsection{Interpreting AI ethics principles}\label{Ch:2:5:interpreting}

Many AI companies have ethics principles \cite{jobin2019global, hagendorff2020ethics}, but “principles alone cannot guarantee ethical AI” \cite{mittelstadt2019principles}. They are necessarily vague and need to be put into practice \cite{morley2020initial, zhou2022ai, seger2022defence}.

\paragraph{Options.} The ethics board could interpret principles in the abstract (e.g. defining terms or clarifying the purpose of specific principles) or in concrete cases (e.g. whether a new research project violates a specific principle). In doing so, it could influence a wide range of risk-related decisions. For example, the board might decide that releasing a model that can easily be misused would violate the principle “be socially beneficial”, which is part of Google’s AI principles (Google, n.d.). When interpreting principles, the board could take a risk-based approach: the higher the risk, the more the company needs to do to mitigate it \cite{baldwin2016driving, mahler2021between, chamberlain2022risk}. The board could also suggest amendments to the principles.

\paragraph{Discussion.} How much would this reduce risk? It will be more effective if the principles play a key role within the company. For example, Google’s motto “don’t be evil”—which it quietly removed in 2018—used to be part of its code of conduct and, reportedly, had a significant influence on its culture \cite{crofts2020negotiating}. Employees could threaten to leave the company or engage in other forms of activism if principles are violated \cite{belfield2020activism}. Interpreting ethics principles would also be more effective if the board’s interpretation is binding (Section \ref{Ch:5:2:binding}), and if the principles are public, mainly because civil society could hold the company accountable \cite{cihon2021corporate}. It would be less effective if the principles are mainly a PR tool. This practice is called “ethics washing” \cite{bietti2020ethics, seger2022defence, floridi2021translating}.

\subsection{Contact point for whistleblowers}\label{Ch:2:6:contact}

Detecting misconduct is often difficult: it is hard to observe from the outside, while insiders might not report it because they face a conflict between personal values and loyalty \cite{jubb1999whistleblowing, dungan2015psychology} or because they fear negative consequences \cite{bjorkelo2013workplace}. For example, an engineer might find a severe safety flaw, but the research lead wants to release the model nonetheless and threatens to fire the engineer if they speak up. In such cases, whistleblower protection is vital.

\paragraph{Options.} An ethics board could protect whistleblowers by providing a trusted contact point. The ethics board could report the case to the board of directors, especially the board risk committee, who could engage with management to do something about it. It could also advise the whistleblower on steps they could take to protect themselves (e.g. seeking legal assistance) or to do something about the misconduct (e.g. leaking the information to the press or a government agency).

\paragraph{Discussion.} The ethics board would be more trustworthy than other organizational units (at least if it is independent from management). But since it would still be part of the company (Section \ref{Ch:3:2:internal}), or at least in a contractual relationship with it (Section \ref{Ch:3:1:external}), confidentiality would be less of a problem. This can be particularly important if the information is highly sensitive and its dissemination could be harmful in itself \cite{urbina2022dual, bostrom2011information, bostrom2019vulnerable}. 

The ethics board can only serve this role if employees trust the ethics board, they know about the board’s commitment to whistleblower protection, and at least one board member needs to have relevant expertise and experience. For more information on the drivers of effective whistleblowing, we refer to the relevant literature \cite{near1995effective, apaza2011makes}. Anecdotally, whistleblowing within large AI companies has had some successes, though it did not always work \cite{cihon2021corporate}. Overall, this role seems very promising, but the issue is highly delicate and could easily make things worse.

\section{Structure}\label{Ch:3:structure}

What should the board’s (legal) structure be? We can distinguish between internal (Section \ref{Ch:3:1:external}) and external structures (Section \ref{Ch:3:2:internal}). The board could also have substructures (Section \ref{Ch:3:3:substructures}).

\subsection{External boards}\label{Ch:3:1:external}

The ethics board could be external. The company and the ethics board could be two separate legal entities. The relationship between the two entities would then be governed by a contract.

\begin{figure}
    \includegraphics[width=\textwidth]{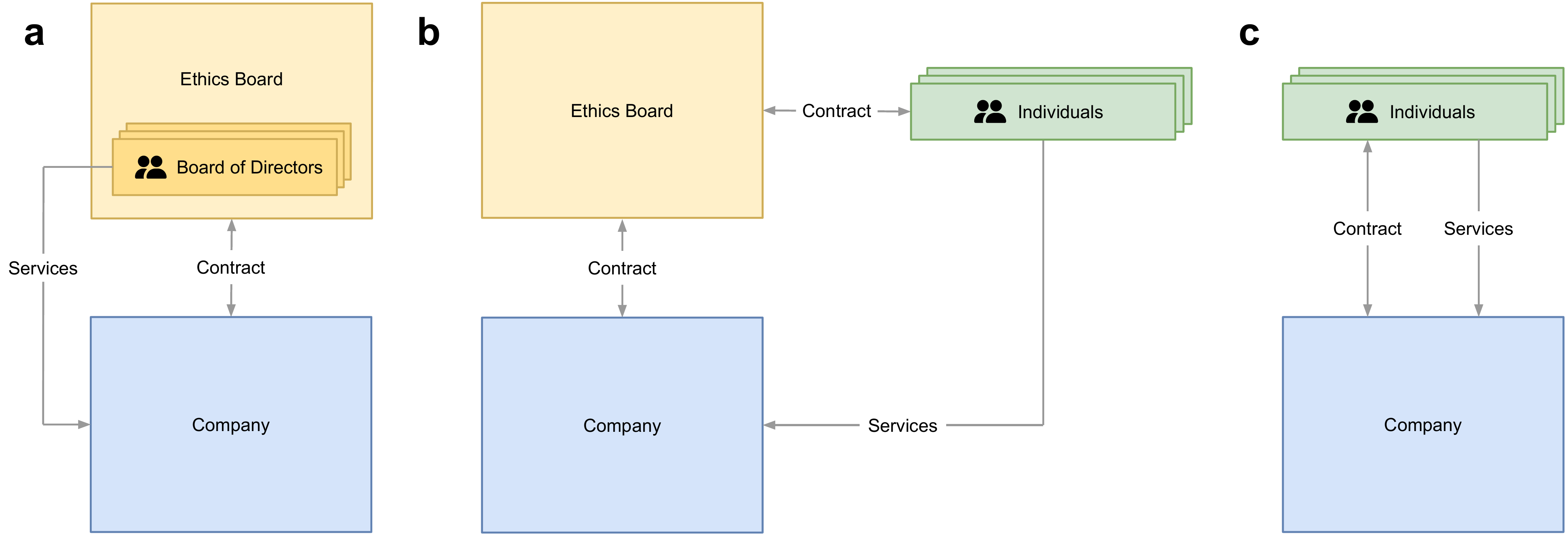}
    \caption{Three potential structures of an external ethics board}
    \label{fig1}
\end{figure}

\begin{figure}
    \includegraphics[width=\textwidth]{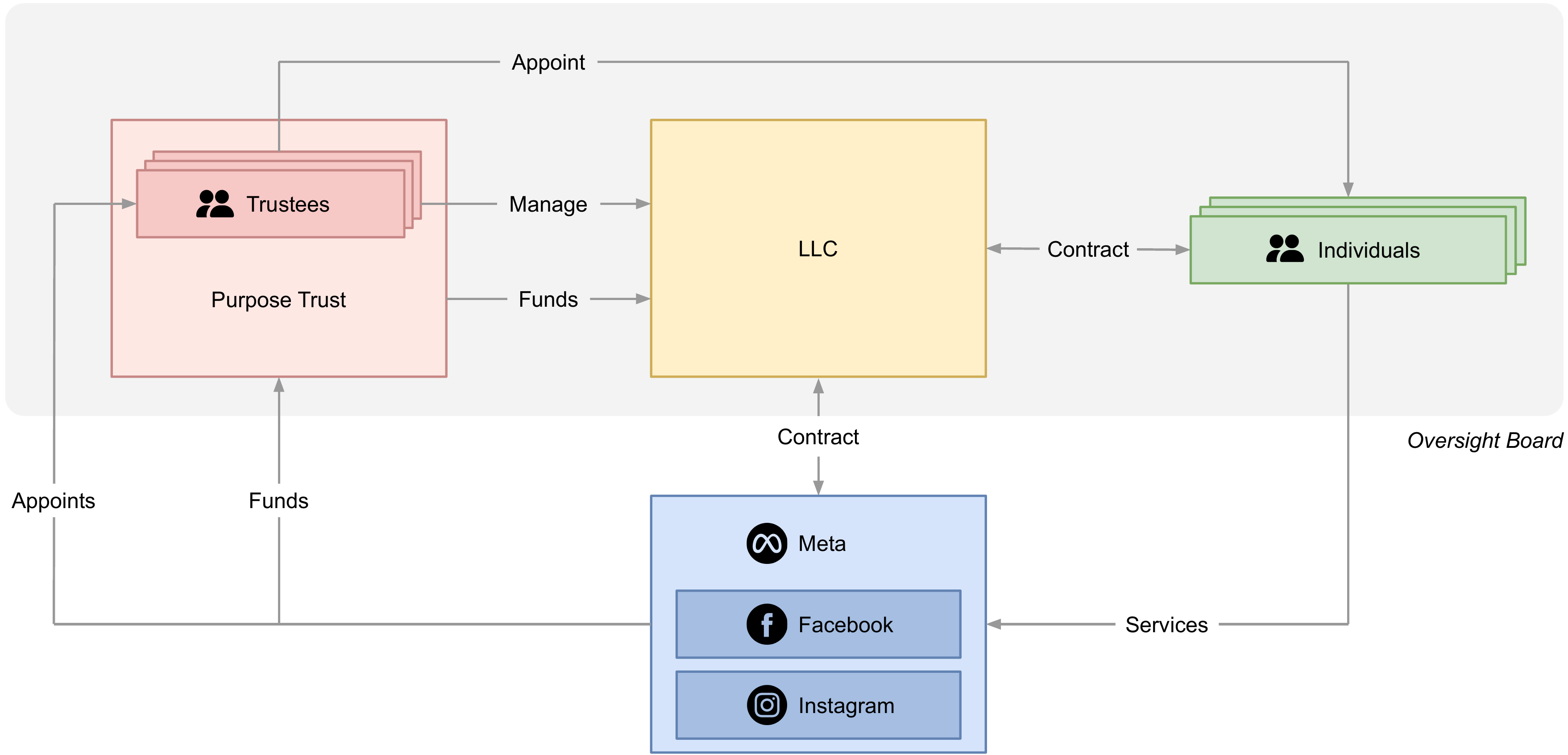}
    \caption{Structure of Meta’s Oversight Board}
    \label{fig2}
\end{figure}

\paragraph{Options.} The ethics board could be a nonprofit organization (e.g. a 501(c)(3)) or a for-profit company (e.g. a public-benefit corporation [PBC]). The individuals who provide services to the company could be members of the board of directors of the ethics board (Figure \ref{fig1}a). Alternatively, they could be a group of individuals contracted by the ethics board (Figure \ref{fig1}b) or by the company (Figure \ref{fig1}c). There could also be more complex structures. For example, Meta’s Oversight Board consists of two separate entities: a purpose trust and a limited liability company (LLC) \cite{oversightboard2023trustees, thomas2019independence}. The purpose trust is funded by Meta and funds the LLC. The trustees are appointed by Meta, appoint individuals, and manage the LLC. The individuals are contracted by the LLC and provide services to Facebook and Instagram (Figure \ref{fig2}).

\paragraph{Discussion.} External ethics boards have a number of advantages: (1) They can legally bind the company through the contractual relationship (Section \ref{Ch:5:1:process}). This would be much more difficult for internal structures (Section \ref{Ch:3:2:internal}). (2) The board would be more independent, mainly because it would be less affected by internal incentives (e.g. board members could prioritize the public interest over the company’s interests). (3) It would be a more credible commitment because it would be more effective and more independent. The company might therefore be perceived as being more responsible. (4) The ethics board could potentially contract with more than one company. In doing so, it might build up more expertise and benefit from economies of scale. But external boards also have disadvantages. We expect that few companies are willing to make such a strong commitment, precisely because it would undermine its independence. It might also take longer to get the necessary information and a nuanced view of the inner workings of the company (e.g. norms and culture).

\subsection{Internal boards}\label{Ch:3:2:internal}

The ethics board could also be part of the company. Its members would be company employees. And the company would have full control over the board’s structure, its activities, and its members.

\paragraph{Options.} An internal board could be a team, i.e. a permanent group of employees with a specific area of responsibility. But it could also be a working group or committee, i.e. a temporary group of employees with a specific area of responsibility, usually in addition to their main activity. For example, DeepMind’s IRC seems to be a committee, not a team \cite{kavukcuoglu2022alphafold, deepmind2023humanrightspolicy}.

\paragraph{Discussion.} The key advantage of internal boards is that it is easier for them to get information (e.g. because they have a better network within the organization). They will typically also have a better understanding of the inner workings of the company (e.g. norms and culture). But internal structures also have disadvantages. They can be disbanded at the discretion of senior management or the board of directors. It would be much harder to play an adversarial role and openly talk about risks, especially when potential mitigations are in conflict with other objectives (e.g. profits). The board would not have much (legal) power. Decisions cannot be enforced. To have influence, it relies on good relationships with management (if collaborative) or the board of directors (if adversarial). Finally, board members would be less protected from repercussions if they advocate for unfavorable measures.

\subsection{Substructures}\label{Ch:3:3:substructures}

Both internal and external boards could have substructures. Certain responsibilities could be delegated to a part of the ethics board.

\paragraph{Options.} Two common substructures are committees and liaisons. (Note that an internal ethics board can be a committee of the company, but the ethics board can also have committees.) (1) Committees could be permanent (for recurring responsibilities) or temporary (to address one-time issues). For example, the board could have a permanent “deployment committee” that reviews model releases (Section \ref{Ch:2:2:overseeing}), or it could have a temporary committee for advising the board on an upcoming M\&A transaction. For more information about the merits of committees in the context of the board of directors, we refer to the relevant literature. Meta’s Oversight Board has two types of committees: a “case selection committee” which sets criteria for cases that the board will select for review, and a “membership committee” which proposes new board members and recommends the removal or renewal of existing members \cite{oversight2022bylaws}. They can also set up other committees.

Liaisons are another type of substructure. Some members of the ethics board could join specific teams or other organizational structures (e.g. attend meetings of research projects or the board of directors). They would get more information about the inner workings of the company and can build better relationships with internal stakeholders (which can be vital if the board wants to protect whistleblowers, see Section \ref{Ch:2:6:contact}). Inversely, non-board members could be invited to attend board meetings. This could be important if the board lacks the necessary competence to make a certain decision (Section \ref{Ch:4:4characteristics:}). For example, they could invite someone from the technical safety team to help them interpret the results of a third-party model audit. Microsoft’s AETHER Committee regularly invites engineers to working groups \cite{microsoft2020principles}.

\paragraph{Discussion.} On the one hand, substructures can make the board more complex and add friction. On the other hand, they allow for faster decision-making because less people are involved and group discussions tend to be more efficient. Against this background, we expect that substructures are probably only needed in larger ethics boards (Section \ref{Ch:4:3:size}).

\section{Membership}\label{Ch:4:membership}

Who should sit on the board? In particular, how should members join (Section \ref{Ch:4:1:joining}) and leave the board (Section \ref{Ch:4:2:leaving})? How many members should the board have (Section \ref{Ch:4:3:size})? What characteristics should they have (Section \ref{Ch:4:4characteristics:})? How much time should they spend on the board (Section \ref{Ch:4:5:time})? And should they be compensated (Section \ref{Ch:4:6:compensation})?

\subsection{Joining the board}\label{Ch:4:1:joining}

How should members join the board?

\paragraph{Options.} We need to distinguish between the appointment of the initial and subsequent board members. Initial members could be directly appointed by the company’s board of directors. But the company could also set up a special formation committee which appoints the initial board members. The former was the case at Axon’s AI and Policing Technologies Ethics Board \cite{axon2019policing}, the latter at Meta’s Oversight Board \cite{oversight2023charter}. Subsequent board members are usually appointed by the board itself. Meta’s Oversight Board has a special committee that selects subsequent members after a review of the candidates’ qualifications and a background check \cite{oversight2023charter}. But they could also be appointed by the company’s board of directors. Candidates could be suggested (not appointed) by other board members, the board of directors, or the general public. At Meta’s Oversight Board, new members can be suggested by other board members, the board of directors, and the general public \cite{oversight2023charter}.

\paragraph{Discussion.} The appointment of initial board members is particularly important. If the company does not get this right, it could threaten the survival of the entire board. For example, Google appointed two controversial members to the initial board which sparked internal petitions to remove them and contributed to the board’s failure \cite{piper2019googlefail}. The appointment should be done by someone with enough time and expertise. This suggests that a formation committee will often be advisable. The board would be more independent if it can appoint subsequent members itself. Otherwise, the company could influence the direction of the ethics board over time.

\subsection{Leaving the board}\label{Ch:4:2:leaving}

How should members leave the board?

\paragraph{Options.} There are at least three ways in which members could leave the board. First, their term could expire. The board’s charter or bylaws could specify a term limit. Members would leave the board when their term expires. For example, at Meta’s Oversight Board, the term ends after three years, but appointments can be renewed twice \cite{oversight2023charter}. Second, members could resign voluntarily. While members might resign for personal reasons, a resignation can also be used to express protest. For example, in the case of Google’s ATEAC, Alessandro Acquisti announced his resignation on Twitter to express protest against the setup of the board \cite{acquisti2019tweet}. Similarly, in the case of Axon’s AI and Policing Technologies Ethics Board, nine out of eleven members publically resigned after Axon announced plans to develop taser-equipped drones to be used in schools without consulting the board first \cite{policing2022resigning}. Third, board members could be removed involuntarily.

\paragraph{Discussion.} Since any removal of board members is a serious step, it should only be possible under special conditions. In particular, it should require a special majority and a special reason (e.g. a violation of the board’s code of conduct or charter). To preserve the independence of the board, it should not be possible to remove board members for substantive decisions they have made.

\subsection{Size of the board}\label{Ch:4:3:size}

How many members should the board have?

\paragraph{Options.} In theory, the board can have any number of members. In practice, most boards have between 10-20 members (Table \ref{table1}).

\paragraph{Discussion.} On the one hand, larger boards can work on more cases and they can go into more detail. They can also be more diverse \cite{gupta2020ai}. On the other hand, it will often be difficult to find enough qualified people. Group discussions in smaller boards tend to be easier and it is easier to reach consensus (e.g. if a qualified majority is required [Section \ref{Ch:5:1:process}]). Smaller boards allow for closer personal relationships between board members. But conflicts of interest could have an outsized effect in smaller boards. As a rule of thumb, the number of members should scale with the board’s workload (“more cases, more members”).

\renewcommand{\arraystretch}{1.4}
\begin{table}
  \caption{Size of different AI ethics boards}
  \label{table1}
  \centering
  \begin{tabular}{lll}
    \toprule
        \textbf{Ethics board}                               & \textbf{Members}  & \textbf{Source} \\
    \midrule
        Meta’s Oversight Board                              & 22                & \cite{oversightboard2023members}\\
        Microsoft’s AETHER Committee                        & 20                & \cite{newman2020decisionpoints}\\
        Google’s ATEAC                                      & 8                 & \cite{walker2019external}\\
        Axon’s AI and Policing Technologies Ethics Board    & 11                & \cite{axon2019policing}\\
        Axon’s Ethics \& Equity Advisory Council            & 11 (US), 7 (UK)   & \cite{axon2022ethicsequity}\\
    \bottomrule
  \end{tabular}
\end{table}

\subsection{Characteristics of members}\label{Ch:4:4characteristics:}

What characteristics should board members have?

\paragraph{Options.} When appointing board members, companies should at least consider candidates’ expertise, diversity, seniority, and public perception.

\paragraph{Discussion.} (1) Different boards will require different types of expertise \cite{sandler2019building}. But we expect most boards to benefit from technical, ethical, and legal expertise. (2) Members should be diverse along various dimensions, such as gender, race, and geographical representation \cite{gupta2020ai}. For example, Meta’s Oversight Board has geographic diversity requirements in its bylaws \cite{oversight2022bylaws}. They should adequately represent historically marginalized communities \cite{mohamed2020decolonial, birhane2022power}. Diverse perspectives are particularly important in the context of risk assessment (Section \ref{Ch:2:3:supporting}). For example, this will make it more likely that unprecedented risks are identified. (3) Board members may be more or less senior. By “seniority”, we mean a person’s position of status which typically corresponds to their work experience and is reflected in their title. More senior people tend to have more subject-matter expertise. The board of directors and senior management might also take them more seriously. As a consequence, it might be easier for them to build trust, get information, and influence key decisions. This is particularly important for boards that only advise and are not able to make binding decisions. However, it will often be harder for the company to find senior people. And in many cases, the actual work is done by junior people. (4) Finally, some board members might be “celebrities”. They would add “glamor” to the board, which the company could use for PR reasons. Inversely, appointing highly controversial candidates (e.g. who express sympathy to extreme political views) might put off other candidates and undermine the board’s credibility.

\subsection{Time commitment}\label{Ch:4:5:time}

How much time should members spend on the board?

\paragraph{Options.} Board members could work full-time (around 40 hours per week), part-time (around 15-20 hours per week), or even less (around 1-2 hours per week or as needed). None of the existing (external) boards seem to require full-time work. Members of Meta’s Oversight Board work part-time \cite{klonick2021insights}. And members of Axon’s AI and Policing Technologies Ethics Board only had two official board meetings per year, with ad-hoc contact between these meetings \cite{axon2019policing}.

\paragraph{Discussion.} The more time members spend working on the board, the more they can engage with individual cases. This would be crucial if cases are complex and stakes are high (e.g. if the board supports pre-deployment risk assessments, see Section \ref{Ch:2:3:supporting}). Full-time board members would also get a better understanding of the inner workings of the company. For some responsibilities, the board needs this understanding (e.g. if the board reviews the company’s risk management practices, see Section \ref{Ch:2:4:reviewving}). However, we expect it to be much harder to find qualified candidates who are willing to work full-time because they will likely have existing obligations or other opportunities. This is exacerbated by the fact that the relevant expertise is scarce. And even if a company finds qualified candidates who are willing to work full-time, hiring several full-time members can be a significant expense.

\subsection{Compensation}\label{Ch:4:6:compensation}

Should board members be compensated?

\paragraph{Options.} There are three options. First, serving on the ethics board could be unpaid. Second, board members could get reimbursed for their expenses (e.g. for traveling or for commissioning outside expertise). For example, Axon paid its board members \$5,000 per year, plus a \$5,000 honorarium per attended board meeting, plus travel expenses (AI and Policing Technologies Ethics Board, 2019). Third, board members could be fully compensated, either via a regular salary or honorarium. For example, it has been reported that members of Meta’s Oversight Board are being paid a six-figure salary \cite{klonick2021insights}.

\paragraph{Discussion.} Not compensating board members or only reimbursing their expenses is only reasonable for part-time or light-touch boards. Full-time boards need to be compensated. Otherwise, it will be extremely difficult to find qualified candidates. For a more detailed discussion of how compensation can affect independence, see Section \ref{Ch:6:1:funding}.

\section{Decision-making}\label{Ch:5:decision-making}

\subsection{Decision-making process}\label{Ch:5:1:process}

How should the board make decisions?

\paragraph{Options.} We expect virtually all boards to make decisions by voting. This raises a number of questions:

\begin{itemize}[leftmargin=2em]
    \item \textbf{Majority.} What majority should be necessary to adopt a decision? Boards could vote by absolute majority, i.e. a decision is adopted if it is supported by more than 50\% of votes. For certain types of decisions, the board may also require a qualified majority (e.g. a unanimous vote or a 67\% majority). Alternatively, boards could vote by plurality (or relative majority), i.e. a decision is adopted if it gets more votes than any other but does not receive more than half of all votes cast. The majority could be calculated based on the total number of board members (e.g. if the board has 10 members, 6 votes would constitute a simple majority), or the number of members present (e.g. if 7 members are present, 4 votes would constitute a simple majority). At Meta’s Oversight board, “outcomes will be determined by majority rule, based on the number of members present” \cite{oversight2022bylaws}.
    \item \textbf{Voting rights.} Who should be able to vote? There are three options. First, all board members could have voting rights. Second, only some board members could have voting rights. For example, only members of subcommittees could be able to vote on issues related to that subcommittee. This is the case at Meta’s Oversight Board \cite{oversight2022bylaws}. It would also be conceivable that some members only advise on special issues; they might be less involved in the board’s day-to-day work. These board members, while formally being part of the board, might not have voting rights. Third, non-board members could have (temporary) voting rights. For example, the board could ask external experts to advise on specific issues. These experts could be granted voting rights for this particular issue.
    \item \textbf{Voting power.} A related, but different question is: how much should a vote count? We expect this question to be irrelevant for most boards, as “one person, one vote” is so commonsensical. However, in some cases, boards may want to deviate from this. For example, the board could use quadratic voting, which allows individuals to express the degree of their preferences, rather than just the direction of their preferences \cite{posner2014quadratic, lalley2018quadratic}.
    \item \textbf{Quorum.} What should the minimum number of members necessary to vote be? This is called a “quorum”. In principle, the quorum can be everything between one and all board members, though there might be legal requirements for some external structures. A natural quorum is the number of board members who could constitute a majority (e.g. more than 50\% of board members if a simple majority is sufficient). It is also possible to have a different quorum for different types of decisions. Note that a lack of quorum might make the decision void or voidable.
    \item \textbf{Voting method.} How should the board vote? The most common voting methods are paper ballots, show of hands, postally, or electronically (e.g. using a voting app). According to its bylaws, voting at Meta’s Oversight Board takes place “in-person or electronically” \cite{oversight2022bylaws}.
    \item \textbf{Abstention.} In some cases, board members may want to abstain from a vote (e.g. because they do not feel adequately informed about the issue at hand, are uncertain, or mildly disapprove of the decision, but do not want to actively oppose it). Abstention could always be permitted or prohibited. The board could also allow abstention for some decisions, but not for others. Board members must abstain if they have a conflict of interest. At Meta’s Oversight Board, abstention is only prohibited for one type of decisions, namely for case deliberation \cite{oversight2022bylaws}.
    \item \textbf{Proxy voting.} Some board members may want to ask someone else to vote on their behalf. This is called “proxy voting”. Proxy voting could always be permitted or prohibited. The board could also allow proxy voting under certain circumstances (e.g. in the event of illness), only for certain decisions (e.g. less consequential decisions), or upon request. Meta’s Oversight Board does not allow proxy voting \cite{oversight2022bylaws}.
    \item \textbf{Frequency of board meetings.} How often should the board meet to vote? There are three options. First, the board could meet periodically (e.g. weekly, monthly, quarterly, or annually). Second, the board could meet on an ad hoc basis. Special meetings could be arranged at the board’s discretion, upon request by the company, and/or based on a catalog of special occasions (e.g. prior to the deployment of a new model). Third, the board could do both, i.e. meeting periodically and on an ad hoc basis. Meta’s Oversight Board meets annually and has special board meetings “in emergency or exceptional cases” \cite{oversight2022bylaws}. Google’s ATEAC planned to have four meetings per year \cite{walker2019external}.
    \item \textbf{In-person or remote meetings.} Should board meetings be held in person or remotely? We expect this design choice to be less important than most others, but it is a necessary one nonetheless. At Meta’s Oversight Board, meetings take place in person, though it does allow exceptions “in limited and exceptional circumstances”; its committees meet either in person or remotely \cite{oversight2022bylaws}.
    \item \textbf{Preparation and convocation of board meetings.} How should board meetings be prepared and convened? More precisely, who can convene a board meeting? What is the notice period? How should members be invited? What should the invitation entail? And do members need to indicate if they will attend? At Meta’s Oversight Board, “written notice of periodic and special meetings must specify the date, time, location, and purpose for convening the board. This notice will be provided at least eight weeks in advance for in-person convenings and, unless in case of imminent emergency, at least two days in advance for remote convenings. Members are required to acknowledge receipt of this notice and also indicate their attendance in a timely fashion” \cite{oversight2022bylaws}.
    \item \textbf{Documentation and communication of decisions.} Finally, it needs to be specified how decisions are documented and communicated. More precisely, which decisions should be documented and communicated? What exactly should be documented and communicated? And who should get access to the documentation? At Facebook’s Oversight Board, “minutes will be taken and circulated to board members within one week” \cite{oversight2022bylaws}. It does not publicly release meeting minutes, but has sometimes allowed reporters in their meetings. Google’s ATEAC planned to “publish a report summarizing the discussions” \cite{walker2019external}. Axon’s AI Ethics Board published two annual reports \cite{policingproject2020reports}. In their 2019 report, they also highlight “the importance of public engagement and transparency” \cite{axon2019policing}.
\end{itemize}

\paragraph{Discussion.} Some of these questions might seem like formalities, but they can significantly affect the board’s work. For example, if the necessary majority or the quorum are too high, the board might not be able to adopt certain decisions. This could bias the board towards inaction. Similarly, if the board is not able to convene ad hoc meetings or only upon request by the company, they would not be able to respond adequately to emergencies.

\subsection{Bindingness of decisions}\label{Ch:5:2:binding}

Should the board’s decisions be binding?

\paragraph{Options.} This mainly depends on the board’s structure (Section \ref{Ch:3:structure}). External boards can be set up in a way that their decisions are binding, i.e. enforceable by legal means. Both parties need to contractually agree that the board’s decisions are in fact binding. This agreement could also contain further details about the enforcement of the board’s decisions (e.g. contractual penalties). It is worth noting, however, that the ethics board cannot force the company to follow its decisions. The worst legal consequence for the company is a contractual liability. If the ethics board is part of the company, it is very difficult, if not impossible, to ensure that the board’s decisions are legally binding. If the board is able to make binding decisions, it needs to be specified whether and, if so, under what conditions the company can override them. For example, the contract could give the company’s board of directors the option to override a decision if they achieve the same voting majority as the ethics board. But even if the board’s decisions are not enforceable by legal means, there are non-legal means that can incentivize the company to follow the board’s decision. For example, the board could make its decisions public, which could spark a public outcry. One or more board members could (threaten to) resign, which might lead to negative PR. Employees could also (threaten to) leave the company (e.g. via an open letter), which could be a serious threat, depending how talent-constraint the company is. Finally, shareholders could engage in shareholder activism. In practice, the only ethics board that is able to make binding decisions is Meta’s Oversight Board, which has the power to override content moderation decisions.

\paragraph{Discussion.} Boards that are able to make legally binding decisions are likely more effective, i.e. they are able to achieve their goals to a higher degree (e.g. reducing risks to an acceptable level). They would also be a more credible commitment to safety and ethics. However, we expect that many companies would oppose creating such a powerful ethics board, mainly because it would undermine the company’s power. There might also be legal constraints on how much power the company can transfer to the ethics board.

\section{Resources}\label{Ch:6:resources}

What resources does the board need? In particular, how much funding does the board need and where should the funding come from (Section \ref{Ch:6:1:funding})? How should the board get information (Section \ref{Ch:6:2:information})? And should it have access to outside expertise (Section \ref{Ch:6:3:expertise})?

\subsection{Funding}\label{Ch:6:1:funding}

How much funding does the board need and where should the funding come from?

\paragraph{Options.} The board might need funding to pay its members salaries or reimburse expenses (Section \ref{Ch:4:6:compensation}), to commission outside expertise (e.g. third-party audits or expert consulting), or to organize events (e.g. in-person board meetings). Funding could also allow board members to spend their time on non-administrative tasks. For example, the Policing Project provided staff support, facilitated meetings, conducted research, and drafted reports for Axon’s former AI and Policing Technologies Ethics Board \cite{policingproject2020reports}. How much funding the board needs varies widely—from essentially no funding to tens of millions of dollars. For example, Meta’s Oversight Board has an annual budget of \$20 million \cite{oversight2022funding}. Funding could come from the company (e.g. directly or via a trust) or philanthropists. Other funding sources do not seem plausible (e.g. state funding or research grants).

\paragraph{Discussion.} The board’s independence could be undermined if funding comes directly from the company. The company could use the provision of funds as leverage to make the board take decisions that are more aligned with its interests. A more indirect funding mechanism therefore seems preferable. For example, Meta funds the purpose trust for multiple years in advance \cite{oversight2022funding}.

\subsection{Information}\label{Ch:6:2:information}

How should the board get information?

\paragraph{Options.} What information the board needs is highly context-specific and mainly depend on the board’s responsibilities (Section \ref{Ch:2:responsibilities}). The board’s structure determines what sources of information are available (Section \ref{Ch:3:structure}). While internal boards have access to some information by default, external boards have to rely on public information and information the company decides to share with them. Both internal and external boards might be able to gather additional information themselves (e.g. via formal document requests or informal coffee chats with employees).

\paragraph{Discussion.} Getting information from the company is convenient for the board, but the information might be biased. The company might—intentionally or not—withhold, overemphasize, or misrepresent certain information. The company could also delay the provision of information or present them in a way that makes it difficult for the board to process (e.g. by hiding important information in long documents). To mitigate these risks, the board might prefer gathering information itself. In particular, the board might want to build good relationships with a few trusted employees. While this might be less biased, it would also be more time-consuming. It might also be impossible to get certain first-hand information (e.g. protocols of past meetings of the board of directors). It is worth noting that not all company information is equally biased. For example, while reports by management might be too positive, whistleblower reports might be too negative. The most objective information will likely come from the internal audit team and external assurance providers \cite{schuett2022lines}. In general, there is no single best information source. Boards need to combine multiple sources and cross-check important information.

\subsection{Outside expertise}\label{Ch:6:3:expertise}

Should the board have access to outside expertise?

\paragraph{Options.} There are at least three types of outside expertise the ethics board could harvest. First, it could hire a specialized firm (e.g. a law or consulting firm) to answer questions that are beyond its expertise (e.g. whether the company complies with the NIST AI Risk Management Framework). Second, it could hire an audit firm (e.g. to audit a specific model, the company’s governance, or its own practices). Third, it could build academic partnerships (e.g. to red-team a model).

\paragraph{Discussion.} It might make sense for the ethics board to rely on outside expertise if they have limited expertise or time. They could also use it to get a more objective perspective, as information provided to them by the company can be biased (Section \ref{Ch:6:2:information}). However, the company might use the same sources of outside expertise. For example, if a company is open to a third-party audit, it would commission the audit directly (why would it ask the ethics board to do it on its behalf?). In such cases, the ethics board would merely “double-check” the company’s or the third party’s work. While the added value would be low, the costs could be high (especially for commissioning an external audit or expert consulting).

\section{Conclusion}\label{Ch:7:conclusion}

\paragraph{Summary.} In this paper, we have identified key design choices that AI companies need to make when setting up an ethics board (RQ1). For each of them, we have listed different options and discussed how they would affect the board’s ability to reduce risks from AI (RQ2). Table \ref{table2} contains a summary of the design choices we have covered.

\begin{table}
  \caption{Summary of design choices}
  \label{table2}
  \centering
  \begin{tabularx}{\textwidth}{p{4.5cm} X}
    \toprule
        \textbf{High-level questions} & \textbf{Sub-questions / options} \\
    \midrule
        What responsibilities should the board have? &
            \vspace{-0.7em} \begin{itemize}[leftmargin=*, nosep]
                \item Advising the board of directors
                \item Overseeing model releases and publications
                \item Supporting risk assessments
                \item Reviewing risk management practices
                \item Interpreting Al ethics principles
                \item Serving as a contact for whistleblowers \vspace{-0.2em}
            \end{itemize} \\
        \vspace{-1.4em} What should the board's legal structure be? &
            \begin{itemize}[leftmargin=*, nosep] \vspace{-1em}
                \item The board could be a separate legal entity that contracts with the company (external board)
                \item It could also be part of the company (internal board)
                \item Should it have substructures (e.g. committees)? \vspace{-0.2em}
            \end{itemize} \\
        \vspace{-1.4em} Who should sit on the board? &
            \begin{itemize}[leftmargin=*, nosep] \vspace{-1em}
                \item How should initial and subsequent members be appointed?
                \item How should they leave the board?
                \item How many members should the board have?
                \item What characteristics should they have?
                \item How much time should they spend on the board?
                \item Should they be compensated? \vspace{-0.2em}
            \end{itemize} \\
        \vspace{-1.4em} How should the board make decisions? &
            \begin{itemize}[leftmargin=*, nosep] \vspace{-1em}
                \item What decision-making process should the board use?
                \item Should its decisions be binding? \vspace{-0.2em}
            \end{itemize} \\
        \vspace{-1.4em} What resources does the should the board need? &
            \begin{itemize}[leftmargin=*, nosep] \vspace{-1em}
                \item How much funding does the board need and where should the funding come from?
                \item How should the board get information?
                \item Should the board have access to outside expertise? \vspace{-0.8em}
            \end{itemize} \\
    \bottomrule
  \end{tabularx}
\end{table}

\paragraph{Key claims.} Throughout this paper, we have made four key claims. First, ethics boards can take many different shapes. Most design choices are highly context-specific. It is therefore very difficult to make abstract recommendations. There is no one-size-fits-all. Second, ethics boards should be seen as an additional “layer of defense”. They do not have an original role in the corporate governance of AI companies. They do not serve a function that no other organizational structure serves. Instead, most ethics boards support, complement, or duplicate existing efforts. While this reduces efficiency, an additional safety net seems warranted in high-stakes situations. Third, merely having an ethics board is not sufficient. Most of the value depends on its members and their willingness and ability to pursue its mission. Thus, appointing the right people is crucial. Inversely, there is precedent that appointing the wrong people can threaten the survival of the entire board. Fourth, while some design choices might seem like formalities (e.g. when the board is quorate), they can have a significant impact on the effectiveness of the board (e.g. by slowing down decisions). They should not be taken lightly.

\paragraph{Questions for further research.} The paper left many questions unanswered and more research is needed. In particular, our list of design choices is not comprehensive. For example, we did not address the issue of board oversight. If an ethics board has substantial powers, the board itself also needs adequate oversight. A “meta oversight board”—a central organization that oversees various AI ethics boards—could be a possible solution. Apart from that, our list of potential responsibilities could be extended. For example, the company could grant the ethics board the right to appoint one or more members of its board of directors. The ethics board could also oversee and coordinate responses to model evals. For example, if certain dangerous capabilities are detected, the company may want to contact government and coordinate with other labs to pause capabilities research.

We wish to conclude with a word of caution. Setting up an ethics board is not a silver bullet—“there is no silver bullet” \cite{openai2022lessonslearned}. Instead, it should be seen as yet another mechanism in a portfolio of mechanisms.

\section*{Acknowledgements}

We are grateful for valuable feedback from Christina Barta, Carrick Flynn, Cullen O’Keefe, Virginia Blanton, Andrew Strait, Tim Fist, and Milan Griffes. Anka Reuel worked on the project during the 2022 CHERI Summer Research Program. All remaining errors are our own.

\bibliographystyle{abbrv}
\bibliography{ms}

\end{document}